\documentclass[twocolumn,showpacs,aps,prl,superscriptaddress]{revtex4}
\usepackage{graphicx}
\usepackage{dcolumn}
\usepackage{epsfig}
\usepackage{amsmath}
\RequirePackage{xspace}

\newcommand{\BaBarYear}    {04}
\newcommand{\BaBarNumber}  {013}
\newcommand{\SLACPubNumber} {10387}

\input babarsym

\def\epem  {\ensuremath{e^+e^-}\xspace}
\newcommand\dbline{\noalign{\vskip 0.10truecm\hrule}\noalign{\vskip 2pt}\noalign{\hrule\vskip 0.10truecm}}
\providecommand{\tbline}{\noalign{\vskip 0.05truecm\hrule\vskip0.05truecm}}

\newcommand\etal{{\it et al.}}
\newcommand{\bma}[1]{\boldmath{$#1$}}
\newcommand{\half}{\ensuremath{{1\over2}}}
\newcommand{\pvec}{{\bf p}}
\newcommand{\calB}{\ensuremath{{\cal B}}}
\providecommand{\bfemsix}{${\cal B} (10^{-6})$}

\newcommand{\DE}{\ensuremath{\Delta E}}
\newcommand{\UfourS}{\ensuremath{\Upsilon(4S)}}
\newcommand{\thetaT}{\ensuremath{\theta_{\rm T}}}
\newcommand{\costhr}{\ensuremath{\cos\thetaT}}
\newcommand{\xf}{\ensuremath{{\cal F}}}
\newcommand{\hel}{\ensuremath{{\cal H}}}
\newcommand{\mres}{\ensuremath{m_{\rm res}}}

\newcommand{\omtoppp}{\ensuremath{{\omega\ra\pip\pim\piz}}}

\newcommand{\etagg}{\ensuremath{\eta_{\gaga}}}
\newcommand{\etappp}{\ensuremath{\eta_{3\pi}}}
\newcommand{\etatogg}{\ensuremath{\eta\ra\gaga}}
\newcommand{\etatoppp}{\ensuremath{\eta\ra\pi^+\pi^-\pi^0}}

\newcommand{\etaptoepp}{\ensuremath{\etapr\ra\eta\pip\pim}}
\newcommand{\etapepp}{\ensuremath{\etapr_{\eta\pi\pi}}}
\newcommand{\etaprg}{\ensuremath{\etapr_{\rho\gamma}}}
\newcommand{\etaptorg}{\ensuremath{\etapr\ra\rho^0\gamma}}

\newcommand{\fetaeta}{\ensuremath{\eta\eta}}
\newcommand{\etaeta}{\ensuremath{\Bz\ra\fetaeta}}

\newcommand{\Betaeta}{\ensuremath{\calB(\etaeta)}}

\newcommand{\uletaeta}{\ensuremath{2.8}}

\newcommand{\fetaggetagg}{\ensuremath{\eta_{\gamma\gamma}\eta_{\gamma\gamma}}}
\newcommand{\fetaggetappp}{\ensuremath{\eta_{\gamma\gamma}\eta_{3\pi}}}
\newcommand{\fetapppetappp}{\ensuremath{\eta_{3\pi}\eta_{3\pi}}}

\newcommand{\fetaphi}{\ensuremath{\eta\phi}}

\newcommand{\Betaphi}{\ensuremath{\calB(\etaphi)}}
\newcommand{\etaphi}{\ensuremath{\Bz\ra\fetaphi}}

\newcommand{\uletaphi}{\ensuremath{1.0}}

\newcommand{\fetaggphi}{\ensuremath{\eta_{\gamma\gamma}\phi}}
\newcommand{\fetapppphi}{\ensuremath{\eta_{3\pi}\phi}}

\newcommand{\retaomega}{\ensuremath{4.0^{+1.3}_{-1.2}\pm0.4}}
\newcommand{\fetaomega}{\ensuremath{\eta\omega}}
\newcommand{\etaomega}{\ensuremath{\Bz\ra\fetaomega}}
\newcommand{\Betaomega}{\ensuremath{\calB(\etaomega)}}
\newcommand{\Retaomega}{\ensuremath{(\retaomega)\times 10^{-6}}}

\newcommand{\fetaggomega}{\ensuremath{\eta_{\gamma\gamma}\omega}}
\newcommand{\fetapppomega}{\ensuremath{\eta_{3\pi}\omega}}
\newcommand{\etapppomega}{\ensuremath{\Bz\ra\fetapppomega}}

\newcommand{\retaetap}{\ensuremath{0.6^{+2.1}_{-1.7}\pm 1.1}}
\newcommand{\fetaetap}{\ensuremath{\eta\etapr}}
\newcommand{\etaetap}{\ensuremath{\Bz\ra\fetaetap}}
\newcommand{\Betaetap}{\ensuremath{\calB(\etaetap)}}

\newcommand{\uletaetap}{\ensuremath{4.6}}

\newcommand{\fetaggetapepp}{\ensuremath{\eta_{\gamma\gamma}\etapr_{\eta\pi\pi}}}
\newcommand{\fetapppetapepp}{\ensuremath{\eta_{3\pi}\etapr_{\eta\pi\pi}}}
\newcommand{\fetaggetaprg}{\ensuremath{\eta_{\gamma\gamma}\etapr_{\rho\gamma}}}
\newcommand{\fetapppetaprg}{\ensuremath{\eta_{3\pi}\etapr_{\rho\gamma}}}

\newcommand{\fetapetap}{\ensuremath{\etapr\etapr}}
\newcommand{\etapetap}{\ensuremath{\Bz\ra\fetapetap}}
\newcommand{\Betapetap}{\ensuremath{\calB(\etapetap)}}

\newcommand{\uletapetap}{\ensuremath{10}}

\newcommand{\fetapeppetapepp}{\ensuremath{\etapr_{\eta\pi\pi}\etapr_{\eta\pi\pi}}}
\newcommand{\fetapeppetaprg}{\ensuremath{\etapr_{\eta\pi\pi}\etapr_{\rho\gamma}}}

\newcommand{\fetapomega}{\ensuremath{\etapr \omega}}
\newcommand{\etapomega}{\ensuremath{\Bz\ra\fetapomega}}
\newcommand{\Betapomega}{\ensuremath{\calB(\etapomega)}}

\newcommand{\uletapomega}{\ensuremath{2.8}}

\newcommand{\fetapeppomega}{\ensuremath{\etapr_{\eta\pi\pi} \omega}}
\newcommand{\fetaprgomega}{\ensuremath{\etapr_{\rho\gamma} \omega}}

\newcommand{\fetapphi}{\ensuremath{\etapr\phi}}
\newcommand{\etapphi}{\ensuremath{\Bz\ra\fetapphi}}
\newcommand{\Betapphi}{\ensuremath{\calB(\etapphi)}}

\newcommand{\uletapphi}{\ensuremath{4.5}}

\newcommand{\fetapeppphi}{\ensuremath{\etapr_{\eta\pi\pi} \phi}}
\newcommand{\fetaprgphi}{\ensuremath{\etapr_{\rho\gamma} \phi}}

\newcommand{\fphiphi}{\ensuremath{\phi\phi}}
\newcommand{\phiphi}{\ensuremath{\Bz\ra\fphiphi}}
\newcommand{\Bphiphi}{\ensuremath{\calB(\phiphi)}}

\newcommand{\ulphiphi}{\ensuremath{1.5}}

\newcommand{\fetapkz}{\ensuremath{\etapr K^0_S}}
\newcommand{\etapKz}{\ensuremath{\Bz\ra\fetapkz}}
\newcommand{\fphikz}{\ensuremath{\phi K^0_S}}
\newcommand{\phiKz}{\ensuremath{\Bz\ra\fphikz}}

\newcommand{\phitoKpKm}{\ensuremath{\phi\ra\Kp\Km}}

\newcommand{\etaprhogam}{\ensuremath{\etapr\ra\rho\gamma}}
\newcommand{\fkstargam}{\ensuremath{K^* \gamma}}
\newcommand{\Bkstargam}{\ensuremath{\Bz\ra\fkstargam}}
\begin{document}


\preprint{\babar-PUB-\BaBarYear/\BaBarNumber} 
\preprint{SLAC-PUB-\SLACPubNumber} 

\begin{flushleft}
\babar-PUB-\BaBarYear/\BaBarNumber \\
 SLAC-PUB-\SLACPubNumber\\
\end{flushleft}

\title{
 \large  \bf\boldmath Searches for \Bz\ Decays to Combinations of Two  Charmless Isoscalar Mesons 
 
}
%
\author{B.~Aubert}
\author{R.~Barate}
\author{D.~Boutigny}
\author{F.~Couderc}
\author{J.-M.~Gaillard}
\author{A.~Hicheur}
\author{Y.~Karyotakis}
\author{J.~P.~Lees}
\author{V.~Tisserand}
\author{A.~Zghiche}
\affiliation{Laboratoire de Physique des Particules, F-74941 Annecy-le-Vieux, France }
% ...
\collaboration{The \babar\ Collaboration}
\noaffiliation

\begin{abstract}
We search for $B$ meson decays into two-body combinations of $\eta,\
\etapr,\ \omega$, and $\phi$ mesons from 89 million \BB\ pairs
 collected with the 
\babar\ detector at the \pep2 asymmetric-energy \epem collider at SLAC.
 We find the branching fraction $\Betaomega = \Retaomega$ with a significance of 4.3~$\sigma$.  
For the other decay modes we set the following 90\%\ confidence level
  upper limits on the branching fractions, in units of $10^{-6}$:
 $\Betaeta <\uletaeta$,  $\Betaetap <\uletaetap$,
$\Betapetap <\uletapetap$, $\Betapomega <\uletapomega$,  $\Betaphi <\uletaphi$,  $\Betapphi
<\uletapphi$, and $\Bphiphi <\ulphiphi$.
\end{abstract}


\maketitle

We report the results of searches for \Bz\ meson decays to two charmless
pseudoscalar mesons \fetaeta, \fetaetap, \fetapetap, to the
pseudoscalar-vector combinations \fetaomega, \fetapomega, \fetaphi,
\fetapphi, and to the vector meson pair \fphiphi.  These together with
$\omega\omega$ and $\omega\phi$ constitute
all combinations involving isospin singlet members of the ground
state pseudoscalar and vector-meson nonets.  These decay modes
 have not been observed
previously; the published experimental upper limits on their branching
fractions lie in the range $(9-60)\times10^{-6}$ \cite{CLEO}.

The all-neutral-meson final states studied here are described
theoretically by  suppressed amplitudes, with predicted branching fractions less
than a few per million by most estimates
\cite{SU3,CHIANGeta,CHIANGVP,GROSS,ALI,LEPAGE,BENEKE,EILAM}.  By bringing the
experimental sensitivity down to this level we can test and constrain
the models.  In particular, these branching fractions or limits bear on the accuracy
with which \CP-violating asymmetry measurements can be interpreted.

Theoretical approaches include those based on flavor SU(3) relations
among many modes \cite{SU3,CHIANGeta,CHIANGVP}, effective Hamiltonians with
factorization and specific $B$-to-light-meson form factors \cite{ALI},
perturbative QCD \cite{LEPAGE}, and QCD factorization \cite{BENEKE}.
The decays to combinations of $\eta^{(\prime)}$ and $\omega$ involve
color-suppressed tree, CKM-suppressed penguin, and flavor-singlet
penguin amplitudes, while only the last of these contributes to those
with a single $\phi$ meson.  The $\Bz
\rightarrow \phi \phi$ decay is a pure penguin annihilation process with an
expected branching fraction of order $10^{-9}$ in the Standard Model
\cite{EILAM}; this mode would therefore be particularly sensitive to
physics beyond the Standard Model.

In the time evolution of \etapKz\ and \phiKz\ a sinusoidal term arises
from interference between decays with and without mixing.
The coefficient $S$ of this term is related to the CKM
phase $\beta = \arg{(-V_{cd} V^*_{cb}/ V_{td} V^*_{tb})}$ if these
decays are dominated by the single amplitude expected in the Standard
Model.
Additional higher-order amplitudes with different weak phases would lead
to deviations $\Delta S$ between  the value 
measured in these rare modes and  the precise determination in the
more copious charmonium \KS\ decays.  
Flavor SU(3) \cite{CHIANGeta,GROSS}\ relates the strength of such
additional amplitudes to the decay rates of two-body \Bz\ decays to
final states containing \piz, $\eta$, and $\etapr$.
The $\eta^{(\prime)}$ combinations reported here provide the strongest constraints.

The results presented here are based on data collected
with the \babar\ detector~\cite{BABARNIM}
at the PEP-II asymmetric $e^+e^-$ collider~\cite{pep}
located at the Stanford Linear Accelerator Center.  An integrated
luminosity of 81.9~fb$^{-1}$, corresponding to 
$N_{\BB}= 88.9 \pm 1.0$ million \BB\ pairs, was recorded at the $\Upsilon (4S)$ resonance (center-of-mass energy $\sqrt{s}=10.58\ \gev$).
A 9.6~fb$^{-1}$  off-resonance data sample, with a center-of-mass energy 40 \mev below the  $\Upsilon (4S)$ resonance, is used to study background
contributions resulting from  \epem $ \rightarrow  q \bar q$ ($q = u$,
$d$, $s$, or $c$) continuum events.

Charged particles from \epem\ interactions are detected, and their
momenta measured, by a combination of a vertex tracker consisting
of five layers of double-sided silicon microstrip detectors, and a
40-layer central drift chamber, both operating in the 1.5-T magnetic
field of a superconducting solenoid. We identify photons and electrons 
using a CsI(Tl) electromagnetic calorimeter.
Further charged-particle identification is provided by the average energy
loss (\dedx ) in the tracking devices and by an internally reflecting
ring-imaging Cherenkov detector (DIRC) covering the central region.

The event selection criteria have been established with studies of
off-resonance data and
simulated Monte Carlo (MC) \cite{geant}\ events of the target decay modes, \BB, and  continuum.  
We select $\eta$, $\etapr$, $\omega$, and $\phi$  candidates through the 
decays \etatogg\ (\etagg), \etatoppp\ (\etappp), 
\etaptoepp\ with \etatogg\ (\etapepp), \etaptorg\ (\etaprg),
\omtoppp,  and \phitoKpKm.  The photon energy $E_{\gamma}$ must be greater than 50 \mev  for
$\pi^0$ and $\eta$ candidates, and  greater than  200 \mev  in \etaprhogam.
We make the following requirements on the invariant mass (in \mev):
 $490< m_{\gaga}<600$ for \etagg, $120< m_{\gaga}<150$ for $\pi^0$,
 $510<m_{\pi\pi}< 1070$ for $\rho^0$,
$520<m_{\pi\pi\pi}< 570$ for
\etappp, $910<(m_{\eta\pi\pi},m_{\rho\gamma})<1000$ for \etapr,
$735<m_{\pi\pi\pi}<825$ for $\omega$, 
and $1009 < m_{K^+K^-} < 1029$ for $\phi$.  We make requirements on DIRC measurements and \dedx to identify pions and 
kaons. Secondary tracks in \etappp, \etapr, and $\omega$ candidates
 must be identified as pions, and in $\phi$ candidates as kaons.
  
A $B$-meson candidate is characterized kinematically by the energy-substituted mass $\mes=\lbrack{(\half s+\pvec_0\cdot\pvec_B)^2/E_0^2-\pvec_B^2}\rbrack^\half$
and energy difference $\DE = E_B^*-\half\sqrt{s}$, where the subscripts $0$ and
$B$ refer to the initial \UfourS\ and to the $B$ candidate, respectively,
and the asterisk denotes the \UfourS\ rest frame. 

Backgrounds arise primarily from random track combinations in $\epem\ra\qqbar$ events. We reject these by using the angle
\thetaT\ between the thrust axis of the $B$ candidate in the \UfourS\
frame and that of the rest of the event.
The distribution of $|\costhr|$ is
sharply peaked near $1.0$ for combinations drawn from jet-like \qqbar\
pairs, and is nearly uniform for $\UfourS \rightarrow \BB$ events.  We require
$|\costhr|<0.9$. To discriminate against $\tau$-pair and two-photon
backgrounds  
we require the event to contain at least the number of charged tracks in the decay mode plus one. For \fetaggetagg\
we require at least 3 charged tracks in the event.

The decay mode \phiphi\ is  very clean. Resolutions on \mes\
and \DE\ are $3.0\ \mev $ and $13.1 \mev$, respectively.  We
define the signal region with cuts of $\pm3 \sigma $
in  \DE\ and $\pm4 \sigma$ in \mes. The number of \phiphi\ candidates in this
signal region is $4.0^{+3.2}_{-1.9}$.
The only source of background is the continuum, estimated with
on-resonance data sidebands to contribute $2.7 \pm 0.4$ events. 

We obtain yields in all other decay modes from unbinned extended maximum-likelihood
(ML) fits.  The principal input observables are \DE\ and \mes. Where  relevant, 
the invariant masses \mres\ 
of the intermediate resonances, a Fisher discriminant \xf, 
and angular variables \hel\ are used.
For $\etagg$,
$\mathcal{H_{\eta}}$ is defined as the cosine of the angle between 
the direction of a daughter $\gamma$ and the flight direction of the
$\eta$ relative to its parent in the $\eta$ rest frame; for \etaprg,
$\mathcal{H_{\rho}}$ is the cosine of the angle between the direction of 
a $\rho$ daughter and the flight direction of the \etapr\ in the $\rho$ rest frame;
for $\omega$, $\mathcal{H_{\omega}}$ is the cosine of the  angle 
in the $\omega$ rest frame between the normal to the $\omega$ decay plane 
and the \Bz\ flight direction. The Fisher discriminant \xf\  combines four variables: the angles with respect to the beam axis of the $B$ momentum and $B$ thrust axis 
(in the \UfourS\ frame), and the zeroth and second angular moments $L_{0,2}$ 
of the energy flow about the \Bz\ thrust axis.  The moments are defined by
$ L_j = \sum_i p_i\times\left|\cos\theta_i\right|^j$,
where $\theta_i$ is the angle with respect to the $B$ thrust axis of
track or neutral cluster $i$, $p_i$ is its momentum, and the sum
excludes the $B$ candidate.
Further cuts on discriminating variables and the  set of probability density 
functions (PDF) used in 
ML fits, specific to each  decay mode, are determined on the basis of studies
with MC samples.
For \fetaggetaprg\  the requirement $|\mathcal{H_{\eta}}| < 0.86$ is used 
to reduce significantly the background from the decay \Bkstargam .
In other decays containing $\eta_{\gamma\gamma}$  we require $|\mathcal{H_{\eta}}| < 0.9$ to remove random combinations with soft photons.
In \fetaggomega\  we apply a cut on the maximum $\gamma$ energy in the center of mass system 
($<2.4$ \gev ) to suppress cross-feed from other \BB\ decays with energetic photons,
and a $\pi^0$ veto to suppress potential cross-feed from $\omega \pi^0$. 

We estimate \BB\ backgrounds using simulated samples of $B$ decays.
The branching fractions in the simulation are based on measured 
values or theoretical predictions. The estimated \BB\ background is
 negligible. 

For each event $i$ and hypothesis $j$ (signal or continuum background),
the likelihood function is
\begin{equation}
{\cal L}= \frac{e^{-\left(\sum n_j\right)}}{N!} \prod_{i=1}^N \left[\sum_{j=1}^m n_j  {\cal P}_j ({\bf x}_i)\right]\ ,
\end{equation}
where  $N$ is the number of input events,  $n_j$ is the number of events for hypothesis $j$ and     ${\cal P}_j ({\bf x}_i)$ the corresponding PDF, evaluated  
with the observables ${\bf x}_i$ of the $i$th event. 
Since the correlations among the observables in the data are small,
we take each  ${\cal P}$  as the product of the PDFs for the separate variables.
We determine the PDF parameters from simulation for the
signal and from sideband
data ($5.20 < \mes\ <5.27$ \gev; $0.1<|\DE |<0.2$ \gev ) 
   for continuum  background.  We float some of the continuum  PDF parameters 
in the maximum likelihood fit. We parameterize each of the functions ${\cal P}_{\rm sig}(\mes),\ 
{\cal  P}_{\rm sig}(\DE),\ { \cal P}_j(\xf),\ $ and the 
peaking components of ${\cal P}_j(\mres)$ with either a Gaussian, the sum of
two Gaussian distributions, or an asymmetric Gaussian function as required to describe the 
distribution.  Slowly varying distributions (mass, energy  
for combinatoric background and angular variables) are represented by linear or 
quadratic dependencies.
The combinatoric background in \mes
is described by the ARGUS function $x\sqrt{1-x^2}\exp{\left[-\xi(1-x^2)\right]}$,
with $x\equiv2\mes/\sqrt{s}$ and parameter $\xi$.
Large control samples of $B$ decays to charmed final states of similar 
topology are used to verify the simulated resolutions in \DE\ and \mes.
Where the control data samples reveal differences from MC in mass or energy
resolution, we shift or scale the resolution used in the likelihood fits.
The bias in the fit is determined from a large set of simulated experiments, each one with the same number of $q \bar{q}$ and signal events as in data.

\begin{table*}[htp]
\caption{
Signal yield (before fit bias correction), detection
efficiency $\epsilon$, daughter branching fraction product,
significance (including systematic errors), measured branching
fraction \calB, and 90\% C.L. 
upper limits (UL) from this and previous work.
}
\label{tab:results}
\begin{tabular}{lccccccc}
\dbline
Mode& \quad Yield \quad&\quad $\epsilon$ (\%) \quad &\quad
$\prod\calB_i$ (\%) \quad&\quad S($\sigma$) \quad &\quad \bfemsix
\quad&\quad This UL $(10^{-6})$ \quad &\quad Previous UL $(10^{-6})$ \cite{CLEO} \quad \\
\tbline
~~\fetaggetagg &   $-7.5^{+6.9}_{-5.9}$ &$21.6$&$15.5$&$0.0$&$-2.4^{+2.3}_{-2.0}$&\\
~~\fetaggetappp &   $0.6^{+6.8}_{-5.8}$ &$16.9$&$17.9$&$0.1$&$0.4^{+2.5}_{-2.2}$&\\
~~\fetapppetappp &   $-0.1^{+3.5}_{-2.3}$ &$12.3$&$5.1$&$0.0$&$-0.4^{+6.2}_{-4.2}$&\\
\bma{\fetaeta}&                   &  &  &\bma{0.0}&\bma{-0.9^{+1.6}_{-1.4}\pm 0.7} &\bma{~ < 2.8} & $~<18$   \\
\hline
~~\fetaggetapepp  &   $-7.1^{+3.7}_{-2.5}$  &$21.5$&$6.9$&$0.0$&$-2.4^{+2.9}_{-1.8}$& \\
~~\fetaggetaprg  &   $0.6^{+5.9}_{-4.3}$  &$20.2$&$11.6$&$0.2$&$0.5^{+3.4}_{-2.4}$ &\\
~~\fetapppetapepp  &   $4.3^{+4.7}_{-3.6}$  &$13.7$&$4.0$&$1.0$&$8.0^{+10.0}_{-7.3}$& \\
~~\fetapppetaprg  &   $1.9^{+7.7}_{-6.2}$  &$13.8$&$6.7$&$0.3$&$2.5^{+9.1}_{-7.3}$& \\
\bma{\fetaetap} &                   &  &  &\bma{0.3} &  \bma{\retaetap}& \bma{~ < \uletaetap}   & $~<27$ \\
\hline
~~\fetapeppetapepp  &$0.3^{+2.6}_{-1.5}$&$14.1$& $3.1$&$0.1$&$0.2^{+6.8}_{-4.0}$ &  \\
~~\fetapeppetaprg  &$4.0^{+7.3}_{-6.2}$&$12.7$& $10.2$&$0.6$&$3.2^{+6.4}_{-5.5}$ &  \\
\bma{\fetapetap} &                   &  &  &\bma{0.4} &\bma{1.7^{+4.8}_{-3.7}\pm 0.6}& \bma{~ < 10}  & $~<47$  \\
\hline
~~\fetaggomega &  $24.2^{+8.2}_{-7.1}$  &$18.1$&$35.1$&$5.1$&$4.4^{+1.5}_{-1.3}$& \\
~~\fetapppomega &  $2.2^{+9.4}_{-8.2}$  &$12.9$&$20.1$&$0.3$&$0.9^{+4.1}_{-3.6}$& \\
\bma{\fetaomega} &                   &  &  &\bma{4.3} &\bma{4.0^{+1.3}_{-1.2}\pm 0.4}& \bma{~ < 6.2}  & $~<12$  \\
\hline
~~\fetapeppomega &   $-3.9^{+4.9}_{-3.4}$  &$14.5$&$15.6$&$0.0$&$-1.8^{+2.5}_{-1.7}$& \\
~~\fetaprgomega &   $1.1^{+6.1}_{-4.0}$  &$13.5$&$26.3$&$0.2$&$0.4^{+1.9}_{-1.3}$ &\\
\bma{\fetapomega} &                   &  &  &\bma{0.0} &\bma{-0.2^{+1.3}_{-0.9}\pm 0.4}& \bma{~ < 2.8}  & $~<60$  \\
\hline
~~\fetaggphi &   $-10.1^{+5.0}_{-3.9}$  &$29.7$&$19.4$&$0.0$&$-2.0^{+1.0}_{-0.7}$& \\
~~\fetapppphi &   $-2.0^{+2.9}_{-1.6}$  &$20.9$&$11.1$&$0.0$&$-0.9^{+1.4}_{-0.8}$& \\
\bma{\fetaphi} &                   &  &  &\bma{0.0} &\bma{-1.4^{+0.7}_{-0.4}\pm 0.2}& \bma{~ < 1.0}   & $~<9$ \\
\hline
~~\fetapeppphi &   $0.5^{+4.0}_{-3.0}$  &$23.2$&$8.6$&$0.1$&$0.3^{+2.2}_{-1.7}$& \\
~~\fetaprgphi &   $8.0^{+8.1}_{-6.9}$  &$22.0$&$14.5$&$1.2$&$2.8^{+2.9}_{-2.4}$& \\
\bma{\fetapphi} &                   &  &  &\bma{0.8} &\bma{1.5^{+1.8}_{-1.5}\pm 0.4}& \bma{~ < 4.5}  & $~<31$  \\    
\hline
\bma{\fphiphi} & $1.3^{+3.2}_{-1.9}$ &$19.9$  &$24.2$&\bma{0.3}&\bma{0.3^{+0.7}_{-0.4} \pm 0.1} & \bma{~ < 1.5}  & $~<12$  \\

\dbline
\end{tabular}
\vspace{-5mm}
\end{table*}

In Table \ref{tab:results} we show the measured yield, the efficiency, 
and the product of daughter branching fractions for each decay mode. 
The efficiency is calculated as the ratio of the numbers of signal MC events entering into the ML fit to the total generated.  
We compute the branching fractions from the fitted  signal event yields, reconstruction efficiency,
daughter branching fractions, and the number of produced $B$ mesons, assuming equal production 
rates of charged and neutral $B$ pairs. We correct the yield for any bias measured with the simulations.
We combine results from different channels by adding the values of $-2\ln{\cal L}$, taking account of the correlated and uncorrelated systematic errors.
We report the statistical significance and the 
branching fractions for the individual decay channels, and for the
combined measurements also the 90\% C.L. upper limits.

The statistical error on the signal yield is taken as the change in 
the central value when the quantity $-2\ln{\cal L}$ increases by one 
unit from its minimum value. The significance is taken as the square root 
of the difference between the value of $-2\ln{\cal L}$ (with systematic 
uncertainties included) for zero signal and the value at its minimum.
The 90\% confidence level (C.L.) upper limit is taken to be the branching 
fraction below which lies 90\% of the total of the likelihood integral 
in the positive branching fraction region.
For the \phiphi\ decay mode the  90\% C.L. upper limit is calculated with the
Feldman-Cousins method \cite{Feld}.

\begin{figure}[!tb]
\vspace{0.5cm}
 \includegraphics[angle=0,width=\linewidth]{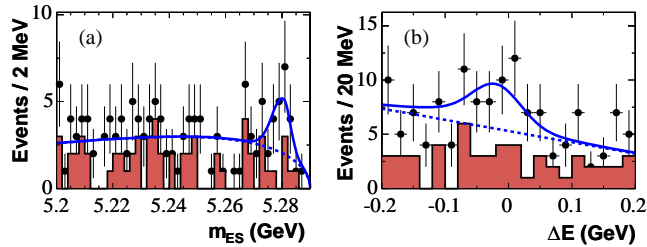}
 \caption{\label{fig:fig01.eps}
Projections of the \Bz\ candidate \mes\ and \DE\ for \etaomega.  
Points with errors represent data, shaded histograms the \etapppomega\ subset, solid curves the full fit functions, and dashed curves the background functions.These plots are made with cuts on probability ratio and thus do not show all events in data samples.}
\end{figure}

In Fig.\ \ref{fig:fig01.eps} we show
 projections onto \mes\ and \DE\ in the analysis of the decays \etaomega. The
histograms show the data after a cut on the probability ratio ${\cal P}_{\rm{sig}}/
({\cal P}_{\rm{sig}}+{\cal P}_{\rm{bkg}})$, where
${\cal P}_{\rm{sig}}$ and ${\cal P}_{\rm{bkg}}$
are the signal and the  continuum background PDFs.  The curve
represents a projection of the PDF obtained from a fit in which the
plotted variable was removed.

The main sources of systematic errors include uncertainty in PDF parameterization (1-2 events) and
 ML fit bias (0.5-2 events).  We estimate these errors with simulated
experiments by
varying PDF parameters within their errors and by embedding MC signal events
inside background events simulated from PDFs. The uncertainty on $N_{\BB}$
is  1.1\%.  Published data
\cite{PDG2002}\ provide the uncertainties in the $B$-daughter branching 
fractions (1-4\%). Other  sources of systematic errors are  track reconstruction
efficiency (1-3\%) and  neutral reconstruction efficiency (5-10\%).
The validity of the fit procedure and PDF parameterization, including
the effects of unmodeled correlations among observables, is checked with simulated 
experiments. The value of the likelihood function found in data is consistent 
with the likelihood distribution found in simulated experiments.  

In the \phiphi\ decay mode the total systematic error is 7.6\%, which we
obtain by adding in quadrature 
the errors due to the different selection cuts, branching fractions of daughters, \Bz\ production, and statistics of the Monte Carlo samples. 

In Grossman \etal\ \cite{GROSS}, $\Delta S = S-\stwob$ for $\Bz\ra\etapr\KS$ is proportional (Eq.\ 10) to the absolute value of a parameter
$\xi_{{\eta^\prime}{K_S}}$ defined in their Eq.\ 8.  A bound
$\left|\xi_{{\eta^\prime}{K_S}}\right|<0.36$ is extracted via Eq.\ 18
from previously measured \Bz\ branching ratios to two-body combinations
of \piz, $\eta$, and \etapr.  The present data improve this limit:
$\left|\xi_{{\eta^\prime}{K_S}}\right|<0.17$.

In conclusion, we have searched for eight \Bz\ decays to charmless isoscalar  meson
pairs.  We obtain  evidence for \etaomega, with a branching fraction 
 $\Betaomega = \Retaomega$ with  4.3~$\sigma$ significance. For the 
other modes our results represent substantial improvements on the previous upper
limits \cite{CLEO}.

We thank Michael Gronau, Yuval Grossman, and Helen Quinn for useful discussions.
We are grateful for the excellent luminosity and machine conditions
provided by our \pep2\ colleagues, 
and for the substantial dedicated effort from
the computing organizations that support \babar.
The collaborating institutions wish to thank 
SLAC for its support and kind hospitality. 
This work is supported by
DOE
and NSF (USA),
NSERC (Canada),
IHEP (China),
CEA and
CNRS-IN2P3
(France),
BMBF and DFG
(Germany),
INFN (Italy),
FOM (The Netherlands),
NFR (Norway),
MIST (Russia), and
PPARC (United Kingdom). 
Individuals have received support from the 
A.~P.~Sloan Foundation, 
Research Corporation,
and Alexander von Humboldt Foundation.

\end{document}